\documentclass[sigconf]{acmart}

\AtBeginDocument{%
  \providecommand\BibTeX{{%
    \normalfont B\kern-0.5em{\scshape i\kern-0.25em b}\kern-0.8em\TeX}}}

\copyrightyear{2021}
\acmYear{2021}
\setcopyright{rightsretained}
\acmConference[AIES '21]{Proceedings of the 2021 AAAI/ACM Conference on AI, Ethics, and Society}{May 19--21, 2021}{Virtual Event, USA}
\acmBooktitle{Proceedings of the 2021 AAAI/ACM Conference on AI, Ethics, and Society (AIES '21), May 19--21, 2021, Virtual Event, USA}\acmDOI{10.1145/3461702.3462471}
\acmISBN{978-1-4503-8473-5/21/05}

\begin{document}
\fancyhead{}

\title{The Coloniality of Data Work in Latin America}

\author{Julian Posada}
\email{julian.posada@mail.utoronto.ca}
\orcid{0000-0002-3285-6503}
\affiliation{%
  \institution{University of Toronto}
  \country{Canada}
}

\begin{abstract}
  This presentation for the AIES ’21 doctoral consortium examines the Latin American crowdsourcing market through a decolonial lens. This research is based on the analysis of the web traffic of ninety-three platforms, interviews with Venezuelan data workers of four platforms, and analysis of the documentation issued by these organizations. The findings show that (1) centuries-old global divisions of labor persist, in this case, with requesters located in advanced economies and workers in the Global South. (2) That the platforms’ configuration of the labor process constrains the agency of these workers when producing annotations. And, (3) that ideologies originating from the Global North serve to legitimize and reinforce this global labor market configuration.
\end{abstract}

\ccsdesc[500]{Human-centered computing~Collaborative and social computing}
\ccsdesc[500]{Information systems~Crowdsourcing}
\ccsdesc[300]{Human-centered computing~Empirical studies in HCI}

\keywords{Coloniality, crowdsourcing, labor, platform}

\maketitle

Firms and research organizations require humans to annotate data to make it compatible with machine learning algorithms \cite{Casilli2019,Gray2019}. These tasks are often outsourced to individuals worldwide through crowdsourcing platforms or infrastructures that serve as marketplaces where human labor is exchanged as a commodity \cite{Casilli2019a}. The firms that operate them consider data workers “independent contractors” without the social and economic benefits and protections of traditional employment relations \cite{Woodcock2020}. The global market of crowdsourcing spans different countries and geographies \cite{Posada2020a}. However, due to the invisible nature of this type of work and the current intricate global data supply chains, understanding the users and configurations of these platforms remains a challenge.

This work presents an analysis of the web traffic from ninety-three crowdsourcing platforms collected during Summer 2020, interviews with data workers from four major platforms, and documentation from these platforms using a decolonial lens. This theory uses a historical perspective to study present power relations that shape society politically, economically, and ideologically \cite{Quijano1992,Mignolo2007}. The analysis of this data suggests the continuation of long historical patterns of domination in how the crowdsourcing market is configured from two levels. A geographical analysis of the web traffic from the platforms shows the continuation of a north-south divide in the distribution of work present in other forms of online work such as freelancing, where the demand for labor comes mainly from advanced economies and the supply from countries in the Global South. However, a unique development in this distribution is the emergence of many workers from Venezuela in the market, a country currently experiencing a severe political and economic crisis.

The analysis of the documentation and the workers’ interviews suggests that platforms constraint their judgment and their labor process. These intermediaries compel them to reproduce the categorization of datasets according to the ideological preferences of requesters, even if they do not always align with the worldviews of data workers. These findings show a continuation of exploitative supply chains in the current artificial intelligence market, where wealthy companies and research institutions in advanced economies profit from the economic and political situations of developing countries to access cheap labor with little regulation from local governments. 

From an ideological perspective, the design of crowdsourcing platforms and their configuration of the labor process evidence a continuation of the suppression of indigenous knowledge by those in power positions and the imposition of their worldviews to individuals from exploited communities. Furthermore, these relations of production are legitimized by ideas and ideologies originated from the Global North, represented for example in the imaginaries of the “entrepreneur” and the “freelancer,” that serve to foment the exploitative working conditions of the market. These configurations continue centuries-long exploitative relations: they are detrimental for both the development of nations and communities in the global south and the pluralistic and ethical development of artificial intelligence systems.

\begin{acks}
Funded by the International Development Research Centre of Canada through a Doctoral Research Award.
\end{acks}

\bibliographystyle{ACM-Reference-Format}
\bibliography{library.bib}


\begin{thebibliography}{7}


\ifx \showCODEN    \undefined \def \showCODEN     #1{\unskip}     \fi
\ifx \showDOI      \undefined \def \showDOI       #1{#1}\fi
\ifx \showISBNx    \undefined \def \showISBNx     #1{\unskip}     \fi
\ifx \showISBNxiii \undefined \def \showISBNxiii  #1{\unskip}     \fi
\ifx \showISSN     \undefined \def \showISSN      #1{\unskip}     \fi
\ifx \showLCCN     \undefined \def \showLCCN      #1{\unskip}     \fi
\ifx \shownote     \undefined \def \shownote      #1{#1}          \fi
\ifx \showarticletitle \undefined \def \showarticletitle #1{#1}   \fi
\ifx \showURL      \undefined \def \showURL       {\relax}        \fi
\providecommand\bibfield[2]{#2}
\providecommand\bibinfo[2]{#2}
\providecommand\natexlab[1]{#1}
\providecommand\showeprint[2][]{arXiv:#2}

\bibitem[\protect\citeauthoryear{Casilli}{Casilli}{2019}]%
        {Casilli2019a}
\bibfield{author}{\bibinfo{person}{Antonio~A. Casilli}.}
  \bibinfo{year}{2019}\natexlab{}.
\newblock \bibinfo{booktitle}{\emph{{En attendant les robots}}}.
\newblock \bibinfo{publisher}{{\'{E}}ditions du Seuil},
  \bibinfo{address}{Paris}. 401 pages.
\newblock
\showISBNx{9782021401882}


\bibitem[\protect\citeauthoryear{Casilli and Posada}{Casilli and
  Posada}{2019}]%
        {Casilli2019}
\bibfield{author}{\bibinfo{person}{Antonio~A. Casilli} {and}
  \bibinfo{person}{Julian Posada}.} \bibinfo{year}{2019}\natexlab{}.
\newblock \showarticletitle{{The Platformisation of Labor and Society}}.
\newblock In \bibinfo{booktitle}{\emph{Society and the Internet}
  (\bibinfo{edition}{vol. 2} ed.)}, \bibfield{editor}{\bibinfo{person}{Mark
  Graham} {and} \bibinfo{person}{William~H. Dutton}} (Eds.).
  \bibinfo{publisher}{Oxford University Press}, \bibinfo{address}{Oxford}.
\newblock


\bibitem[\protect\citeauthoryear{Gray and Suri}{Gray and Suri}{2019}]%
        {Gray2019}
\bibfield{author}{\bibinfo{person}{Mary~L. Gray} {and}
  \bibinfo{person}{Siddharth Suri}.} \bibinfo{year}{2019}\natexlab{}.
\newblock \bibinfo{booktitle}{\emph{{Ghost Work: How to Stop Silicon Valley
  from Building a New Global Underclass}}}.
\newblock \bibinfo{publisher}{Houghton Mifflin Harcourt},
  \bibinfo{address}{Boston, MA}. 288 pages.
\newblock


\bibitem[\protect\citeauthoryear{Mignolo}{Mignolo}{2007}]%
        {Mignolo2007}
\bibfield{author}{\bibinfo{person}{Walter~D. Mignolo}.}
  \bibinfo{year}{2007}\natexlab{}.
\newblock \showarticletitle{{Introduction: Coloniality of power and de-colonial
  thinking}}.
\newblock \bibinfo{journal}{\emph{Cultural Studies}} \bibinfo{volume}{21},
  \bibinfo{number}{2-3} (\bibinfo{date}{mar} \bibinfo{year}{2007}),
  \bibinfo{pages}{155--167}.
\newblock
\showISSN{0950-2386}
\urldef\tempurl%
\url{https://doi.org/10.1080/09502380601162498}
\showDOI{\tempurl}


\bibitem[\protect\citeauthoryear{Posada}{Posada}{2020}]%
        {Posada2020a}
\bibfield{author}{\bibinfo{person}{Julian Posada}.}
  \bibinfo{year}{2020}\natexlab{}.
\newblock \showarticletitle{{The Future of Work Is Here: Toward a Comprehensive
  Approach to Artificial Intelligence and Labour}}.
\newblock \bibinfo{journal}{\emph{Ethics in Context}} \bibinfo{number}{56}
  (\bibinfo{year}{2020}).
\newblock
\showISSN{10185909}


\bibitem[\protect\citeauthoryear{Quijano}{Quijano}{1992}]%
        {Quijano1992}
\bibfield{author}{\bibinfo{person}{An{\'{i}}bal Quijano}.}
  \bibinfo{year}{1992}\natexlab{}.
\newblock \showarticletitle{{Colonialidad y modernidad/racionalidad}}.
\newblock \bibinfo{journal}{\emph{Per{\'{u}} Ind{\'{i}}gena}}
  \bibinfo{volume}{13}, \bibinfo{number}{29} (\bibinfo{year}{1992}),
  \bibinfo{pages}{11--20}.
\newblock


\bibitem[\protect\citeauthoryear{Woodcock and Graham}{Woodcock and
  Graham}{2020}]%
        {Woodcock2020}
\bibfield{author}{\bibinfo{person}{Jamie Woodcock} {and} \bibinfo{person}{Mark
  Graham}.} \bibinfo{year}{2020}\natexlab{}.
\newblock \bibinfo{booktitle}{\emph{{The Gig Economy: A Critical
  Introduction}}}.
\newblock \bibinfo{publisher}{Polity Press}, \bibinfo{address}{London}. 160
  pages.
\newblock


\end{thebibliography}

\end{document}